\documentclass{emulateapj}

\newcommand{\philtrans} {Phil.\ Trans.\ Roy.\ Soc.\ Lond.}
\newcommand{\naturwiss} {Naturwiss.}
\newcommand{\za}        {Zeitschrift f.\ Astrophys.}
\newcommand{\an}        {Astronomische Nachrichten}

\shorttitle{Coronal heating through braiding of magnetic field lines}
\shortauthors{Peter, Gudiksen \& Nordlund}

\begin{document}


\title{Coronal heating through braiding of magnetic field lines}

\author{Hardi Peter\altaffilmark{1}}
\affil{Kiepenheuer-Institut f\"ur Sonnenphysik, 79104 Freiburg, Germany;
       peter@kis.uni-freiburg.de}

\author{Boris V. Gudiksen\altaffilmark{2}}
\affil{Institute of Theoretical Astrophysics, University of Oslo, Norway}

\author{{\AA}ke Nordlund\altaffilmark{3}}
\affil{Astronomical Observatory, NBIfAFG, University of Copenhagen, Denmark}

\begin{abstract}
%
Cool stars like our Sun are surrounded by a million degree hot outer
atmosphere, the corona. Since more than 60 years the physical nature of
the processes heating the corona to temperatures well in excess of those
on the stellar surface remain puzzling. Recent progress in observational
techniques and numerical modeling now opens a new window to approach this
problem.
We present the first coronal emission line spectra synthesized from
three-dimensional numerical models describing the evolution of the
dynamics and energetics as well as of the magnetic field in the corona.
In these models the corona is heated through motions on the stellar
surface that lead to a braiding of magnetic field lines inducing currents
which are finally dissipated.
These forward models enable us to synthesize observed properties like
(average) emission line Doppler shifts or emission measures in the outer
atmosphere, which until now have not been understood theoretically, even
though many suggestions have been made in the past.
As our model passes these observational tests, we conclude that the flux
braiding mechanism is a prime candidate for being the dominant
heating process of the magnetically closed corona of the Sun and
solar-like stars.
%
\end{abstract}


\section{Introduction}

Shortly after it was realized in the 1930ies that the corona is
hot \citep{Grotrian:1939,Edlen:1943}, first proposals for the heating
mechanism were made based on upward propagating sound waves generated on
the stellar surface \citep{Schwarzschild:1948} because convective motions
contain far more energy than what is needed to heat the corona.
Later it became clear that the heating mechanism has to be related to the
magnetic field dominating in the corona.
The coronal magnetic field is rooted in the photosphere, however, where
it is dominated by the kinetic energy of the convection.
Thus at the surface magnetic field-lines are pushed around, and the
(stochastic) footpoint motions result in a braiding of the magnetic field
lines in the corona.
This implies that magnetic field gradients are built up, inducing currents
which are finally dissipated and thus heat the corona
as suggested by \cite{Parker:1972,Parker:1994} --- see also
\cite*{Sturrock+Uchida:1981,Parker:1983,vanBallegooijen:1986,Heyvaerts+Priest:1992}.
Thus the magnetic field acts as the
agent to channel energy from the cool surface into the hot corona.
Several studies have been carried out to investigate the role of footpoint
motions for coronal heating. 
For example \cite{Priest+Schrijver:1999} studied the influence of moving
magnetic sources of opposite polarity in the photosphere on reconnection 
in the above corona and \cite{Priest+al:2002} constructed a model view for
coronal heating based on flux-tube tectonics related to the ever changing
structure of the magnetic carpet \citep{Schrijver+al:1998}.
Many other mechanisms have been discussed to heat the corona, e.g.\ based
on waves, but there are too many of them to be reviewed here \citep[see
the SOHO 15 proceedings for an up to date overview,][]{Danesy:2004}.

Even though the proposal of field line braiding has been widely discussed,
it was not until recently that numerical modeling made it possible to test
whether this mechanism actually works.
Studies had been conducted for the flux braiding on small scales,
investigating the amount and character of the energy
deposition \citep{Hendrix+al:1996,Galsgaard+Nordlund:1996}.
Finally a small active region could be described in a numerical
experiment, resulting in a realistic looking corona consisting
of a number of loop-like
structures \citep{Gudiksen+Nordlund:2002,Gudiksen+Nordlund:2004a,Gudiksen+Nordlund:2004b}.

\begin{figure*}
\centerline{\includegraphics[width=\textwidth]{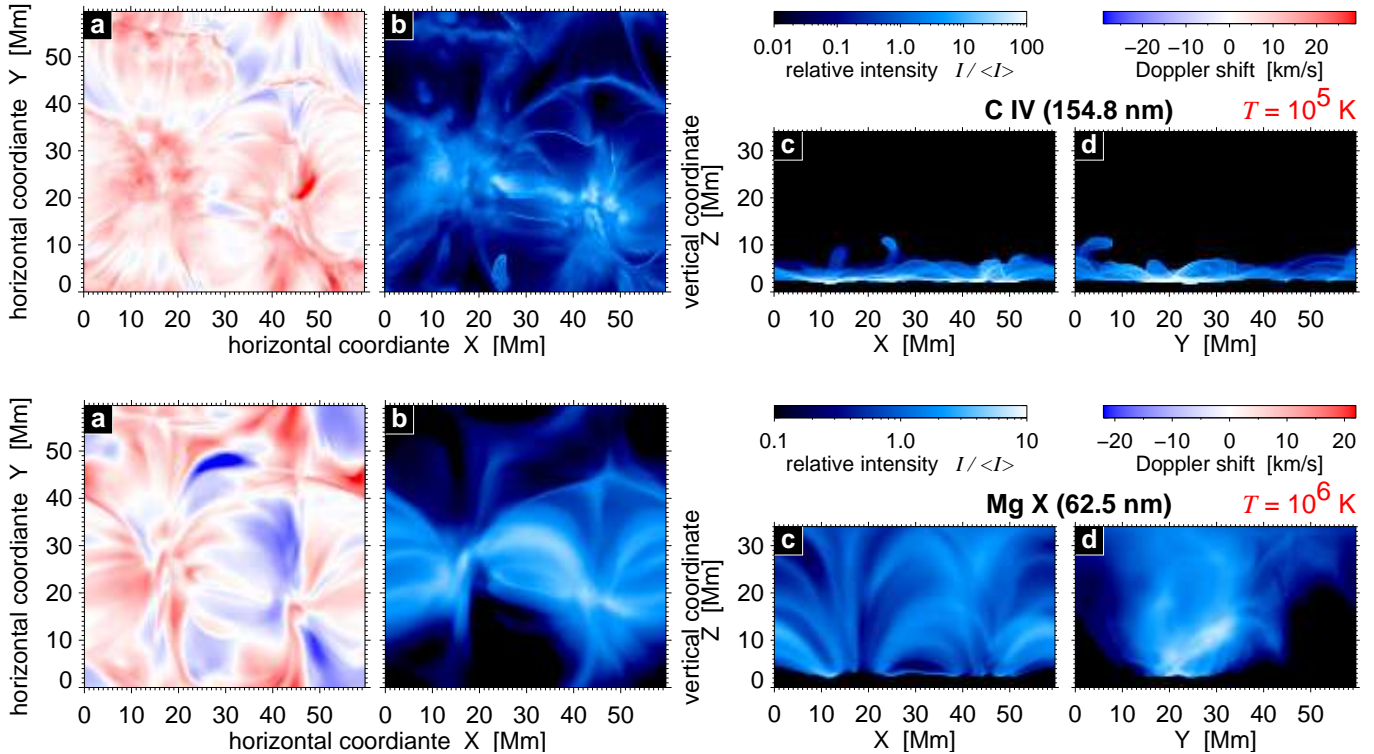}}
\caption{%
Spatial maps in Doppler shift and intensity for the emission lines of
\ion{C}{4} at 154.8\,nm (top row) and \ion{Mg}{10} at 62.5\,nm (bottom
row).
Panels (a) show Doppler shifts of the synthesized spectra as seen from
straight above, panels (b) show the same for line intensity.
This corresponds to the appearance near the center of the stellar disk.
Panels (c) and (d) show side views of the computational box along the x and
y axis in line intensity, corresponding to limb observations.
The intensities $I$ are scaled according to the average intensity
$\langle{I}\rangle$ of the respective map.
Please note the different color tables, especially that the \ion{C}{4}
intensity shows much higher contrast than \ion{Mg}{10}.
Despite the pretty smooth appearance in intensity, the Doppler shift map of
\ion{Mg}{10} is highly structured.
\label{F:images}}
\end{figure*}

\begin{figure}
\centerline{\includegraphics[width=\columnwidth]{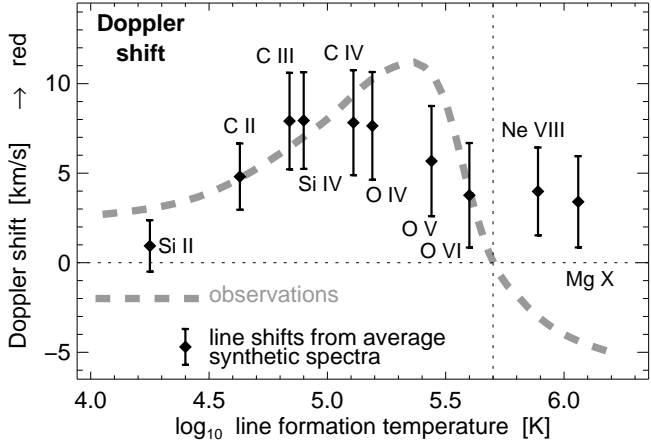}}
\caption{%
Comparison of synthesized and observed Doppler shifts.
The diamonds show the Doppler shifts of the temporally and spatially
averaged spectra for a number of emission lines formed over temperatures
from the transition region and low corona.
The bars indicate the standard deviation (scatter) of the Doppler shifts
in a time-averaged spatial map when looking from straight above at the
computational box (cf.\ panels a in Fig.\ \ref{F:images}).
The thick dashed line shows the trend as found in observations.
\label{F:shift}}
\end{figure}

\begin{figure}
\centerline{\includegraphics{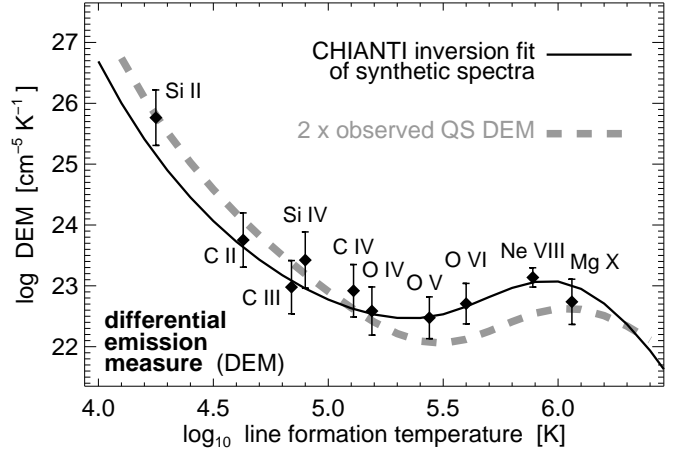}}
\caption{%
Differential emission measure (DEM) as following from
the synthesized line intensities compared to observations.
The solid line shows the fit from the DEM inversion based on the spatially
and temporally averaged synthetic spectra.
The bars indicate the standard deviation of line intensities in a spatial
map.
The thick dashed line is based on a DEM inversion using observed quiet Sun
disk center line radiances from SUMER  \citep{Wilhelm+al:1998}
scaled by a factor of two to match the active region model.
\label{F:dem}}
\end{figure}

In this work we go one step further and synthesize emission line spectra
from such \emph{ab-initio}\/ coronal models (Fig.\ \ref{F:images}) and
compute average properties of these spectra, especially emission measures
and Doppler shifts (Figs.\ \ref{F:shift}, \ref{F:dem}), which  can be
considered as a test for the energetics and the dynamics in the model
corona, respectively.
This allows a robust comparison of the model not only to the Sun but also
to other stars.

\section{MHD model and spectral synthesis}

The solar atmosphere is modeled from the photosphere to the lower corona
using a sixth order fully compressible 3D magneto-hydrodynamics (MHD) code
on a staggered non-equidistant mesh.
The simulated volume is $60\times 60$ Mm$^2$ horizontally and 37 Mm
vertically.
It includes the \cite{Spitzer:1956} heat conductivity along the magnetic
field, and a cooling function representing radiative losses in the
optically thin corona.
In the optically thick photosphere and chromosphere the temperature is
kept near a prescribed temperature profile by a Newtonian temperature
relaxation scheme.
The initial magnetic field is a potential extrapolation of an MDI/SOHO
high resolution magnetogram of AR 9114 scaled to fit into the
computational domain.
The lower boundary is stressed by a time dependent velocity field,
constructed from a Voronoi-tesselation \citep{Okabe+al:1992}, shown to
reproduce the granulation pattern \citep{Schrijver+al:1997}.
The velocity field reproduces the geometrical pattern as well as the
amplitude  power spectrum of the velocity and the vorticity leaving no
free parameters.

The braiding of the magnetic field by the photospheric motions rapidly
produces an intermittent corona in both time and space with a typical
temperature of one million K, during the whole simulated timespan of
${\sim}50$ minutes.
The time and space averaged heating function decreases exponentially with
height, producing a heat input to the corona of
$2{-}8{\times}10^3\,\mbox{W}\,\mbox{m}^{-2}$ in agreement with coronal
energy losses estimated from observations \citep{Withbroe+Noyes:1977}.
The amount of heating produced is  constant or if anything rises as
numerical resolution increases
\citep{Hendrix+al:1996,Galsgaard+Nordlund:1996} and therefore the amount of
heating produced in this simulation is a lower limit.
Thus the spatial resolution of the model (400 km) is not sufficient to
resolve the reconnection process, of course, but based on the above work
the amount of energy input into the corona on larger scales should be of
the right order.
Details on and results from this MHD model can be found in
 \cite{Gudiksen+Nordlund:2004a,Gudiksen+Nordlund:2004b}.


Using density, velocity and temperature from the MHD
model the emissivity for a number of emission lines is evaluated at
each grid point in the computational domain.
By integrating along a line of sight this finally results in maps of
spectra when looking at the computational box from, e.g., straight above.

Almost all emission lines from the low corona and transition region are in
the extreme ultraviolet (EUV; $\approx$500-1500\AA), optically thin and
excited predominantly through electron collisions.
The emissivity $\varepsilon$ is given through the energy $h\nu$, the
Einstein coefficient $A_{21}$ and the upper level population $n_2$ for the
transition from an upper level 2 to a lower level 1, i.e.\ $\varepsilon =
h\nu \, n_2 A_{21}$.
To calculate the upper level density one has to evaluate the balances for
excitation and ionization.
In the latter case we assume ionization equilibrium.
We have checked ionization and advection times and found that for the
present models ionization equilibrium is a reasonable choice.
Furthermore we assume constant abundances with photospheric values.
All these quantities, i.e.\ ionization, excitation, abundances, Einstein
coefficients and transition energies, have been evaluated using the
atomic data package Chianti  \citep{Dere+al:1997,Young+al:2003}, finally
resulting in the emissivity at each grid point depending on the density
and temperature there.
The spectral profiles at each grid point are assumed to be a Gaussian
with a line width corresponding to the thermal width
$(2k_{\rm{B}}T/m)^{1/2}$, where $T$ is the temperature at the respective
grid point and $m$ the mass of the ion emitting the line.
These Gaussians are shifted by the component of the velocity along the line
of sight at the respective grid point.
Finally we integrate the spectra along the line of sight (e.g.\ the
vertical direction) to construct a map of spectra.

These synthesized spectral maps can the be analyzed exactly as it is done
for observations, e.g.\ for a spatial scan using the
SUMER instrument on-board SOHO \citep{Wilhelm+al:1995}.
The resulting spatial maps in line intensity and line shift for a snapshot
of the model run can be seen in panels (a) and (b) of Fig.\ \ref{F:images}
for the lines of \ion{C}{4} at 154.8\,nm and \ion{Mg}{10}
at 62.5\,nm, formed around $10^5$\,K and $10^6$\,K respectively.
The panels (c) and (d) show the intensity images when
viewing the box from the sides, i.e.\ along the x- and y-axis.

\section{Discussion}

The synthesized corona as shown in Fig.\ \ref{F:images} is made up by
numerous individual loops, and the appearance is remarkably different in
the two lines.
At high temperatures the structures are rather washed out and less sharp,
partly reflecting that heat conduction (${\propto}T^{5/2}$) is very 
efficient at high temperatures.
These findings agree well with real observations, and the synthetic
maps roughly resemble maps of solar observations.

Furthermore these images reveal that in the model corona the low
temperature emission originates not only from the low parts of the high
temperature loops, but that a multitude of low lying cool dense loops are
present.
Defining the footpoints of the hot loops by circles (covering about 2\% of
the area) we estimated that only some 10 to 15\% of the emission from
typical transition region lines formed below 200\,000\,K originates from
that area.
Thus the transition region emission is enhanced in the footpoint regions of
hot loops, but the majority is emitted in cooler structures.
This supports earlier sketches of the structure of the low
corona \citep{Dowdy+al:1986,Peter:2001:sec}, but now for the first time this
multi-component or multi-loop structure is based on a solid model of the
coronal structures.

It is also important to note that the Doppler shift images of the
high-temperature line shows much more structures than the intensity image
(Fig.\ \ref{F:images}, panels a and b in lower row), reflecting the highly
dynamic nature of the low corona.
This shows the importance of using EUV spectrographs with sufficient
spectral resolution to learn about the basic coronal processes.


The first major test is provided through average Doppler shifts, which are
shown in Fig.\ \ref{F:shift}.
The temporally and spatially averaged shifts of synthesized spectra for a
number of emission lines are displayed as diamonds.
The bars represent the scatter in the spatial maps (standard deviation;
cf.\ Fig.\ \ref{F:images}).
The underlying thick dashed line shows the observed average variation of
line shifts with temperature compiled from recent quiet Sun
studies \citep{Peter+Judge:1999}.
Active regions \citep{Teriaca+al:1999:ar} and a large number of cool
stars \citep{Wood+al:1997,Pagano+al:2004} show a similar behavior.
The coherence of the average Doppler shifts synthesized from the model and
the observed ones is remarkable.
Also the large scatter is as found with observations \citep{Peter:1999full}.
For the first time the overall variation of Doppler shifts as a function
of line formation temperature could be reproduced --- the line shifts are
caused by flows along the magnetic structures induced by asymetric heating.
Many attempts have been made to understand this curve since the first
recognition of systematic line shifts in the solar transition
region \citep{Doschek+al:1976}.
However, many of these models failed \citep{Peter+Judge:1999} and none of
them reached the level of qualitative and quantitative agreement achieved
by our study.

Only at the highest temperatures does our model not reproduce the
recently observationally established
blueshifts \citep{Peter:1999full,Peter+Judge:1999}.
This could be due to the influence from the impenetrable upper boundary 
condition of the MHD model, which may be expected to quench flows along
magnetic field lines that intersect the upper boundary.
New models, where the upper boundary is shifted higher up into the corona, 
will further reduce the boundary effects and could provide an even better
match to the observed Doppler shifts.


The other major test is to check the emission measure distribution, which
describes the emission efficiency at a given temperature and is defined as
$n^2\,{\rm{d}}h/{\rm{d}}T$ (for electron density $n$ and temperature
gradient with height ${\rm{d}}T/{\rm{d}}h$) \citep{Mariska:1992}.
Using the synthesized lines (averaged spatially and temporally) we have
performed an inversion to obtain the emission measure curve, using the
Chianti package \citep{Young+al:2003}.
The result is shown as a solid line in Fig.\ \ref{F:dem}.
The bars indicate the scatter of the emissivities this inversion is
based upon (standard deviation). 
Over-plotted as a thick dashed line is an inversion from observed quiet
Sun spectra of the same lines from SUMER full disk
data \citep{Wilhelm+al:1998}.
for \ion{Mg}{10} we evaluated disk center
spectra as no full disk scans exist in this line.
The observed quiet Sun emission measure is scaled by a factor of two, to
be comparable to our active region calculation.

The agreement of the emission measure curve from the observations and the
synthetic spectra is remarkable.
Previous (1D or 2D) models failed to reproduce this emission measure curve
at low temperatures by orders of magnitude, basically because they had
not enough emitting material below $10^5$\,K.
It has been proposed, among other suggestions, that in the lowest corona a
large number of cool dense structures might
exist \citep{Gabriel:1976,Dowdy+al:1986}.
This proposition is supported by our results, where enough energy is released
very low in the corona and transition region so that enough dense cool
structures form.
Thus we can unambiguously reproduce the high emission measures at low
temperatures, for the first time without any fine-tuning.
As the emission measure curve is a general feature of a large variety of
cool stars including our Sun, this suggests that the flux-braiding
mechanism may be important in a wide variety of cool stars.

Future work will have to concentrate on improving the coronal MHD model as
well as the spectral synthesis.
This includes the addition of the super-granular structures to the active
region patterns, which is of vital importance to understand the structure
of the quiet solar corona.
Furthermore larger scale shear motions would be needed to achieve a higher
level of activity.
In this case the  plasma motions are expected to be faster, and therefore
requiring the inclusion of non-equilibrium ionization effects when
calculating the emission line spectra.

\section{Conclusions}

This work is the first to reproduce the Doppler shifts and emission measure
distributions in the corona of the Sun and many cool stars in a
quantitative and qualitative way.
Further work will be needed to apply the model also to a wider variety of
stellar activity, especially when thinking of stars with different
magnetic field structures \citep{Donati+al:1999} or convection
patterns \citep{Freytag+al:2002} on the stellar surface.
This will show in which types of cool star coronae this heating mechanism
is the dominant one.
But already we can conclude that flux braiding is a prime candidate for
heating in the magnetically closed parts of the coronae of the
Sun and solar-like stars.


\acknowledgments
Computing time for the MHD-modeling was provided by the Swedish
National Allocations Committee and by the Danish Center for Scientific
Computing.



\begin{thebibliography}{38}
\expandafter\ifx\csname natexlab\endcsname\relax\def\natexlab#1{#1}\fi

\bibitem[{Danesy(2004)}]{Danesy:2004}
Danesy, D., ed. 2004, Coronal Heating --- Proceedings of the SOHO 15
  conference, ESA SP-575 (ESA Publications Division)

\bibitem[{Dere {et~al.}(1997)Dere, Landi, Mason, {Monsignori Fossi}, \&
  Young}]{Dere+al:1997}
Dere, K.~P., Landi, E., Mason, H.~E., {Monsignori Fossi}, B.~C., \& Young,
  P.~R. 1997, \aaps, 125, 149

\bibitem[{Donati {et~al.}(1999)Donati, {Collier Cameron}, Hussain, \&
  Semel}]{Donati+al:1999}
Donati, J.-F., {Collier Cameron}, A., Hussain, G. A.~J., \& Semel, M. 1999,
  \mnras, 302, 437

\bibitem[{Doschek {et~al.}(1976)Doschek, Feldman, \& Bohlin}]{Doschek+al:1976}
Doschek, G.~A., Feldman, U., \& Bohlin, J.~D. 1976, \apj, 205, L177

\bibitem[{Dowdy {et~al.}(1986)Dowdy, Rabin, \& Moore}]{Dowdy+al:1986}
Dowdy, J.~F., Rabin, D., \& Moore, R.~L. 1986, \solphys, 105, 35

\bibitem[{Edl{\'e}n(1943)}]{Edlen:1943}
Edl{\'e}n, B. 1943, \za, 22, 30

\bibitem[{Freytag {et~al.}(2002)Freytag, Steffen, \& Dorch}]{Freytag+al:2002}
Freytag, B., Steffen, M., \& Dorch, B. 2002, \an, 323, 213

\bibitem[{Gabriel(1976)}]{Gabriel:1976}
Gabriel, A.~H. 1976, \philtrans, A\,281, 339

\bibitem[{Galsgaard \& Nordlund(1996)}]{Galsgaard+Nordlund:1996}
Galsgaard, K. \& Nordlund, {\AA}. 1996, \jgr, 101, 13445

\bibitem[{Grotrian(1939)}]{Grotrian:1939}
Grotrian, W. 1939, \naturwiss, 27, 214

\bibitem[{Gudiksen \& Nordlund(2002)}]{Gudiksen+Nordlund:2002}
Gudiksen, B. \& Nordlund, {\AA}. 2002, \apj, 572, L113

\bibitem[{Gudiksen \& Nordlund(2004{\natexlab{a}})}]{Gudiksen+Nordlund:2004b}
---. 2004{\natexlab{a}}, \apj, submitted, astro-ph/0407266

\bibitem[{Gudiksen \& Nordlund(2004{\natexlab{b}})}]{Gudiksen+Nordlund:2004a}
---. 2004{\natexlab{b}}, \apj, submitted, astro-ph/0407267

\bibitem[{Hendrix {et~al.}(1996)Hendrix, {van Hoven}, Mikic, \&
  Schnack}]{Hendrix+al:1996}
Hendrix, D.~L., {van Hoven}, G., Mikic, Z., \& Schnack, D.~D. 1996, \apj, 470,
  1192

\bibitem[{Heyvaerts \& Priest(1992)}]{Heyvaerts+Priest:1992}
Heyvaerts, J. \& Priest, E.~R. 1992, \apj, 390, 297

\bibitem[{Mariska(1992)}]{Mariska:1992}
Mariska, J.~T. 1992, The solar tansition region (Cambridge Univ. Press,
  Cambridge)

\bibitem[{Okabe {et~al.}(1992)Okabe, Boots, \& Sugihara}]{Okabe+al:1992}
Okabe, A., Boots, B., \& Sugihara, K. 1992, Spatial tessellations. Concepts and
  applications of voronoi diagrams (Wiley Series in Probability and
  Mathematical Statistics, Chichester, New York)

\bibitem[{Pagano {et~al.}(2004)Pagano, Linsky, Valenti, \&
  Duncan}]{Pagano+al:2004}
Pagano, I., Linsky, J.~L., Valenti, J., \& Duncan, D.~K. 2004, \aap, 415, 331

\bibitem[{Parker(1972)}]{Parker:1972}
Parker, E.~N. 1972, \apj, 174, 499

\bibitem[{Parker(1983)}]{Parker:1983}
---. 1983, \apj, 264, 642

\bibitem[{Parker(1994)}]{Parker:1994}
---. 1994, Spontaneous current sheets in magnetic fields (Oxford University
  Press)

\bibitem[{Peter(1999)}]{Peter:1999full}
Peter, H. 1999, \apj, 516, 490

\bibitem[{Peter(2001)}]{Peter:2001:sec}
---. 2001, \aap, 374, 1108

\bibitem[{Peter \& Judge(1999)}]{Peter+Judge:1999}
Peter, H. \& Judge, P.~G. 1999, \apj, 522, 1148

\bibitem[{Priest {et~al.}(2002)Priest, Heyvaerts, \& Title}]{Priest+al:2002}
Priest, E.~R., Heyvaerts, J.~F., \& Title, A.~M. 2002, \apj, 576, 533

\bibitem[{Priest \& Schrijver(1999)}]{Priest+Schrijver:1999}
Priest, E.~R. \& Schrijver, C.~J. 1999, \solphys, 190, 1

\bibitem[{Schrijver {et~al.}(1997)Schrijver, Hagenaar, \&
  Title}]{Schrijver+al:1997}
Schrijver, C.~J., Hagenaar, H.~J., \& Title, A.~M. 1997, \apj, 475, 328

\bibitem[{Schrijver {et~al.}(1998)Schrijver, Title, Harvey,
  {et~al.}}]{Schrijver+al:1998}
Schrijver, C.~J., Title, A.~M., Harvey, K.~L., {et~al.} 1998, Nature, 394, 152

\bibitem[{Schwarzschild(1948)}]{Schwarzschild:1948}
Schwarzschild, M. 1948, \apj, 107, 1

\bibitem[{Spitzer(1956)}]{Spitzer:1956}
Spitzer, L. 1956, Physics of fully ionized gases (Interscience, New York)

\bibitem[{Sturrock \& Uchida(1981)}]{Sturrock+Uchida:1981}
Sturrock, P.~A. \& Uchida, Y. 1981, \apj, 246, 331

\bibitem[{Teriaca {et~al.}(1999)Teriaca, Banerjee, \&
  Doyle}]{Teriaca+al:1999:ar}
Teriaca, L., Banerjee, D., \& Doyle, J.~G. 1999, \aap, 349, 636

\bibitem[{{van Ballegooijen}(1986)}]{vanBallegooijen:1986}
{van Ballegooijen}, A.~A. 1986, \apj, 311, 1001

\bibitem[{Wilhelm {et~al.}(1995)Wilhelm, Curdt, Marsch,
  {et~al.}}]{Wilhelm+al:1995}
Wilhelm, K., Curdt, W., Marsch, E., {et~al.} 1995, \solphys, 162, 189

\bibitem[{Wilhelm {et~al.}(1998)Wilhelm, Lemaire, Dammasch,
  {et~al.}}]{Wilhelm+al:1998}
Wilhelm, K., Lemaire, P., Dammasch, I.~E., {et~al.} 1998, \aap, 334, 685

\bibitem[{Withbroe \& Noyes(1977)}]{Withbroe+Noyes:1977}
Withbroe, G.~L. \& Noyes, R.~W. 1977, \araa, 15, 363

\bibitem[{Wood {et~al.}(1997)Wood, Linsky, \& Ayres}]{Wood+al:1997}
Wood, B.~E., Linsky, J.~L., \& Ayres, T.~R. 1997, \apj, 478, 745

\bibitem[{Young {et~al.}(2003)Young, {Del Zanna}, Landi, Dere, Mason, \&
  Landini}]{Young+al:2003}
Young, P.~R., {Del Zanna}, G., Landi, E., Dere, K.~P., Mason, H.~E., \&
  Landini, M. 2003, \apjs, 144, 135

\end{thebibliography}

\end{document}